\newcommand{\CLASSINPUTtoptextmargin}{2.12cm}
\newcommand{\CLASSINPUTbottomtextmargin}{2.12cm}
\documentclass[conference,a4paper]{IEEEtran}
\IEEEquantizetextheight{c}
\IEEEsettextwidth{14mm}{14mm}
\IEEEsetsidemargin{c}{0mm}

\IEEEoverridecommandlockouts
\usepackage{cite}
\usepackage{amsmath,amssymb,amsfonts,mathtools}
\usepackage{algorithmic}
\usepackage{graphicx}
\usepackage{textcomp}
\usepackage{xcolor}
\usepackage{siunitx}
\usepackage{acronym}
\usepackage{todonotes}
\usepackage{microtype}
\usepackage[caption=true,font=footnotesize,labelfont=footnotesize,textfont=footnotesize]{subfig}
\usepackage{multirow}
\usepackage{url}
\usepackage{caption}
\def\BibTeX{{\rm B\kern-.05em{\sc i\kern-.025em b}\kern-.08em
    T\kern-.1667em\lower.7ex\hbox{E}\kern-.125emX}}

\def\ve#1{\ensuremath{{\mathchoice{\mbox{\boldmath$\displaystyle #1$}}	%
			{\mbox{\boldmath$\textstyle #1$}}			%
			{\mbox{\boldmath$\scriptstyle #1$}}			%
			{\mbox{\boldmath$\scriptscriptstyle #1$}}}}}	%
\def\ma#1{\ensuremath{{\mathchoice{\mbox{\boldmath$\displaystyle #1$}}		%
			{\mbox{\boldmath$\textstyle #1$}}				%
			{\mbox{\boldmath$\scriptstyle #1$}}			%
			{\mbox{\boldmath$\scriptscriptstyle #1$}}}}}
\newcommand{\argmin}{\mathop{\mathrm{argmin}}}	

\newcommand{\norm}[1]{\ensuremath{ \Vert #1 \Vert}}
\newcommand{\RR}{\ensuremath{\mathbb{R}}}

\newcommand{\ZZ}{\ensuremath{\mathbb{Z}}}
\newcommand{\degi}[1]{\mathop{\mathrm{d_\mathrm{D}^{-}}}\ensuremath{(#1)}}
\newcommand{\dego}[1]{\mathop{\mathrm{d_\mathrm{D}^{+}}}\ensuremath{(#1)}}
\newcommand{\supp}{\mathop{\mathrm{supp}}}
\newcommand{\tomat}[1]{\mathop{\mathrm{mat(#1)}}}
\newcommand{\shortset}[1]{\mathop{[#1]}}

\newtheorem{definition}{Definition}
\newtheorem{remark}{Remark}

\begin{document}
\acrodef{lcc}[LCC]{linear computation coding}
\acrodef{fs}[FS]{fully sequential}
\acrodef{fp}[FP]{fully parallel}
\acrodef{ua}[MA]{mixed algorithm}
\acrodef{dag}[DAG]{directed acyclic graph}
\acrodef{scm}[SCM]{single constant multiplication}
\acrodef{mcm}[MCM]{multiple constant multiplication}
\acrodef{cmvm}[CMVM]{constant matrix vector multiplication}
\acrodef{dmp}[DMP]{discrete matching pursuit}
\acrodef{rs}[RS]{reduced state}
\acrodef{lz}[LZ]{Lempel-Ziv}
\acrodef{sqnr}[SQNR]{signal to quantization noise ratio}
\acrodef{csd}[CSD]{canonically signed digit}
\acrodef{fpga}[FPGA]{Field Programmable Gate Array}
\acrodef{lut}[LUT]{lookup table}
\acrodef{nn}[NN]{neural network}
\acrodef{ls}[LS]{least squares}

\title{Graph-based Algorithms for Linear Computation Coding\\
\thanks{This work was supported by Deutsche Forschungsgemeinschaft (DFG) under the project Computation Coding (MU-3735-/8-1).}

}

\author{\IEEEauthorblockN{Hans Rosenberger\IEEEauthorrefmark{1}, Ali Bereyhi\IEEEauthorrefmark{2}, Ralf R. M\"uller\IEEEauthorrefmark{1}}
\IEEEauthorblockA{\IEEEauthorrefmark{1}Institute for Digital Communications, Friedrich-Alexander-Universität (FAU), Erlangen, Germany \\
\{hans.rosenberger, ralf.r.mueller\}@fau.de \\
\IEEEauthorrefmark{2}Wireless Computing Lab, University of Toronto, Canada \\
ali.bereyhi@utoronto.ca
}
}

\maketitle

\begin{abstract}
    We revisit existing linear computation coding (LCC) algorithms, and introduce a new framework that measures the computational cost of computing multidimensional linear functions, not only in terms of the number of additions, but also with respect to their suitability for parallel processing. Utilizing directed acyclic graphs, which correspond to signal flow graphs in hardware, we propose a novel LCC algorithm that controls the trade-off between the total number of operations and their parallel executability. Numerical evaluations show that the proposed algorithm, constrained to a fully parallel structure, outperforms existing schemes.
\end{abstract}


\section{Introduction}
Over-parameterized \acp{nn} have achieved many of the recent advancements in improving inference accuracy.
Many real-world applications of these very large \acp{nn} require both real-time inference and operate in a resource constrained environment. 
It is therefore of great importance to implement them with minimal computational complexity.
Various research efforts have been directed towards improving \ac{nn} efficiency, including pruning, knowledge distillation, quantization and \ac{nn}-hardware co-design~\cite{Gholami_2021survey,Neill_2020overview}.

\Ac{lcc} introduces an analytical framework that invokes the idea of sparse matrix decomposition to reduce the computational cost of computing matrix-vector products, i.e. the lossy compression of a multidimensional linear function with constant coefficients.
Earlier studies on \ac{lcc} mainly focus on the number of additions as the metric of computational complexity \cite{M_ller_2022,Muller_2022,Rosenberger:2023,Karataev_2023}. Though important, this metric is not the only concern in many applications. 

In this paper, we revisit the earlier \ac{lcc} studies from a new perspective on computational complexity, in which not only the number of operations, but also their order matters. Our interest follows from a simple fact: optimizing the order in which the operations are carried out enables us to fully exploit the potential of \textit{parallel processing}. We use the notion of a \ac{dag}, closely corresponding to the signal flow graph of a hardware implementation, to 
develop a new \ac{lcc} algorithm. 
The proposed scheme explicitly tunes the structure of the \acs{dag} and outperforms existing algorithms 
on parallel processing units.



\subsection{Notation}
Vectors and matrices are denoted by lower- and upper-case boldface letters, e.g. $\ve{x}$ and $\ma{X}$, respectively.
The Euclidean and Frobenius norms are shown by $\norm{\cdot}_2$ and $\norm{\cdot}_\mathrm{F}$, respectively. 
The matrix transpose is denoted by $(\cdot)^\mathrm{T}$.
The augmented identity matrix with dimension $N \times K$ is denoted by $\ma{I}_{N \times K}$, and the $j$-th row unit vector in $K$ dimensions by $\ve{1}_{j,K}$.
The function $\supp(\ve{x})$ returns the indices in the support of $\ve{x}$, i.e. the set of all indices $i$ where $x_i \neq 0$.

Sets are specified by upper case caligraphic letters, e.g. $\mathcal{A}$. 
We use the notation $|\mathcal{A}|$ to represent the cardinality of $\mathcal{A}$.
A \ac{dag} is denoted by $D = ( \mathcal{C}, \mathcal{A} )$, where $\mathcal{C} \subset \RR^{1 \times K}$ is the ordered set of all vertices and $\mathcal{A}$ the set of arcs (directed edges).
The indegree and outdegree of a vertex $\ve{c} \in \mathcal{C}$ are denoted by $\degi{\ve{c}}$ and $\dego{\ve{c}}$, respectively.
Given a \ac{dag} $D = (\mathcal{C}, \mathcal{A})$ and a vertex $\ve{c} \in \mathcal{C}$, $\mu_\mathrm{D}(\ve{c})$ denotes the depth of $\ve{c}$, i.e. the longest path from any node $\ve{c}' \in \mathcal{C}$ 
to node $\ve{c}$.
The operator $\tomat{\cdot}$ converts a vertex set $\mathcal{C} = \{ \ve{c}_1, \dots, \ve{c}_L \} \subset \RR^{1 \times K}$ with $|\mathcal{C}|=L$ to its corresponding matrix, i.e. $\ma{C} = \tomat{\mathcal{C}} = \left[ \ve{c}_1, \dots, \ve{c}_L \right] \in \RR^{L \times K}$.
Unless otherwise specified, $\ve{c}_i$ denotes the $i$-th element in the set $\mathcal{C}$ or the $i$-th row vector of the corresponding matrix $\ma{C} = \tomat{\mathcal{C}}$.
The notation $[N]$ is an abbreviation for the set $\{1,\dots,N\}$.

\section{Preliminaries}
Consider the matrix vector product
\begin{align}
    \ve{y} = \ma{T} \ve{x}
\end{align}
with the arbitrary, but constant, matrix $\ma{T} \in \RR^{N \times K}$ and the arbitrary input vector $\ve{x} \in \RR^{K \times 1}$.
Our goal is to approximately compute $\ve{y} \in \RR^{N \times 1}$ with minimum effort.
Calculating the matrix-vector product straightforwardly requires $NK$ multiplications and $N(K-1)$ additions.
Using a finite-precision representation of $\ma{T}$, a multiplication can be reduced to additions and bitshifts.
Quantizing the matrix entries independently, it is well known that each additional bit on average improves the \ac{sqnr} by \SI{6}{\decibel} while requiring half an extra addition.
Using the \ac{csd} representation~\cite{Booth_1951}, i.e. allowing for subtractions as well, the \ac{sqnr} even improves by \SI{14.5}{\decibel} per digit.
However, by quantizing the operations of a matrix-vector product jointly, far larger gains are possible~\cite{Aksoy_2015, M_ller_2022}.
\subsection{Addition as a Fundamental Operation}
\begin{definition}[Fundamental Operation]
Let $\mathcal{C} \subset \RR^{1 \times K}$ denote a set of $L$ vectors and be called a codebook.
We define the fundamental operation as the linear combination of at most $S$ vectors contained in $\mathcal{C}$, or, more formally:
\begin{align}
    \mathrm{add}_S(\ve{\omega}_S, \mathcal{C}) &= \ve{\omega}_S \tomat{\mathcal{C}}
\end{align}
with  $\ve{\omega}_S \in \mathcal{W}_S$, where
\begin{align}
    \mathcal{W}_S = &\biggl\{ \ve{\omega} = \sum_{s=1}^S i_s \ve{1}_{j_s, \, L} : i_s \in \mathcal{M} \subseteq \{ 0, \pm 2^\ZZ \}, j_s \in \shortset{L} \; \forall s \biggr\}.
\end{align}
The nonzero coefficients of $\ve{\omega}_S \in \mathcal{W}_S$ are restricted to the set of (sums of) signed powers of two, corresponding only to bitshifts in hardware, which can be considered computationally cheap.\footnote{In this paper we consider the set of wiring coefficients to be unrestricted, i.e. $\mathcal{M} = \{ 0,\pm 2^\ZZ \}$.
For some applications, it is beneficial to restrict the coefficients to a subset.
Efficient strategies for such cases are investigated in~\cite{Karataev_2023}.}
The computational cost of a fundamental operation is governed by the at most $S-1$ additions needed to form the linear combination.
\end{definition}

Given a codebook $\mathcal{C}$ and using the notion of the fundamental operation, our aim is now to approximate a target vector $\ve{t}$ by a single fundamental operation.
We call this objective wiring.
Mathematically we aim to solve the following \ac{ls} problem:
\begin{align}\label{eq:sparse}
    w(\ve{t}, \mathcal{C}, S) = \underset{\ve{\omega}_S \in \mathcal{W}_S}{\argmin} \norm{\ve{t} - \ve{\omega}_S \tomat{\mathcal{C}}}_2,
\end{align}
which can be equivalently seen as a sparse recovery problem~\cite{Foucart2013} due to the restricted support of $\ve{\omega}_S$.

The minimization over the set of discrete vectors $\mathcal{W}_S$ in~\eqref{eq:sparse} is an NP-hard problem.
Hence, an optimal solution is generally computationally intractable.
Therefore, we resort to the following two suboptimal approaches:
\begin{itemize}
    \item \textit{\Ac{dmp}}~\cite{M_ller_2022}: Start with $\ve{\omega} \gets \ve{0}$. Find the vector in $\ve{c}_i \in \mathcal{C}$ scaled by a signed power of two that reduces the error to $\ve{t}$ maximally and update $\ve{\omega}$ in the $i$-th component. Repeat $S$ times. 
    \item \textit{\Ac{rs} approach}~\cite{Rosenberger:2023}: Procedure similar to \ac{dmp}. However, instead of choosing in each iteration the best vector minimizing the error, we retain a list of the $Q$ best linear combinations in each iteration and choose the combination with minimum error at termination. This procedure enables a performance close to full search at a reasonable time complexity~\cite{Rosenberger:2023}.
\end{itemize}

To quantify the ability of a codebook $\mathcal{C}$ to approximate the matrix $\ma{T}$ with row vectors $\ve{t}_n$, we use the \ac{sqnr} defined as
\begin{align}
    \mathrm{SQNR}(\ma{T}, \mathcal{C}) = \frac{\norm{\ma{T}}_\mathrm{F}^2}{\sum_{n=1}^N \norm{\ve{t}_n - w(\ve{t}_n, \mathcal{C}, 1) \tomat{\mathcal{C}}}_2^2}.
\end{align}
Note that $w(\ve{t}_n, \mathcal{C}, 1) \tomat{\mathcal{C}}$ finds the vector in $\mathcal{C}$ scaled by a signed power of two, that approximates $\ve{t}_n$ best.
As $S=1$, this is only a selection and potentially a bitshift, no additions are required.

\subsection{\Acf{cmvm}}
Using the notion of a fundamental operation, any matrix-vector product with finite precision can now be expressed as a \ac{dag} with $K$ input and $N$ output vertices. Input vertices are all vertices with no preceding fundamental operations, i.e. $\{ \ve{c} \in \mathcal{C} | \degi{\ve{c}} = 0 \}$. 
Likewise, output vertices have no arcs directed to subsequent vertices ($\{ \ve{c} \in \mathcal{C} | \dego{\ve{c}} = 0 \}$).
In such a graph, each vertex, except the input vertices, corresponds to one fundamental operation, and each directed arc is labeled with a signed power of two.
An example of such a \ac{dag} is depicted in Fig.~\ref{fig::dagexample}.
It is our goal, given some target matrix $\ma{T}$, to find a \ac{dag} requiring a minimum of computations given some fidelity constraint.
We can therefore define now a \ac{cmvm} problem.

\begin{definition}[\ac{cmvm} Problem]
For all fundamental operations assume without loss of generality $S=2$.
Given a target matrix $\ma{T}$ and some positive parameter $\epsilon$, find a \ac{dag} $D = (\mathcal{C}, \mathcal{A})$ with vertex set $\mathcal{C} \subset \RR^{1 \times K}$, that solves
\begin{subequations}
    \begin{align}
        \min \quad &|\mathcal{C}| \\
        \mathrm{s.t.} \quad &\mathrm{SQNR}(\ma{T},\mathcal{C}) > \epsilon \\
        &\ve{c}_m = \ve{1}_{m,K} \qquad \forall m \in \shortset{K} \\
        &\ve{c}_l = \mathrm{add}_2 (\ve{\omega}_2, \{ \ve{c}_i \in \mathcal{C} : i \in \shortset{l-1} \}) \quad \forall l > K  
    \end{align}
\end{subequations}
\end{definition}
The \ac{cmvm} problem is at least NP-complete.
Similar to \ac{mcm}~\cite{Voronenko_2007}, it is an even broader generalization of the \ac{scm} problem,\footnote{The optimization of the multiplication of a constant scalar to a scalar variable.} which is known to be NP-complete~\cite{Cappello_1984,Garey_1979}. 
Hence, by polynomial reduction the \ac{cmvm} problem has to be at least as difficult.
As the optimal solution is generally computationally intractable, we focus for the remainder of this paper on the development of efficient heuristics for obtaining decomposition \acp{dag}.

\begin{remark}
Throughout the paper we do not specify the set of arcs $\mathcal{A}$ of a \ac{dag} explicitly for reasons of brevity.
As new vertices are created from an initial codebook, i.e. the set of unit vectors, by means of fundamental operations, implicitly $\mathcal{A}$ is defined uniquely\footnote{Uniqueness only refers in that context to the start and endpoint of individual arcs, not their labeling. For example two different fundamental operations, differing in their labeling/bitshift, might produce the same result, i.e. $\ve{c}_2 = \ve{c}_1 - 1/4\ve{c}_1 = 1/2\ve{c}_1 + 1/4\ve{c}_1$.} by $\mathcal{C}$ for any decomposition \ac{dag} $D = (\mathcal{C},\mathcal{A})$ as well.
\end{remark}

\subsection{Computational Cost}\label{sec::compcost}
\begin{figure}
    \centering
    \subfloat[\Ac{dag}]{
    \includegraphics[width=.8\columnwidth]{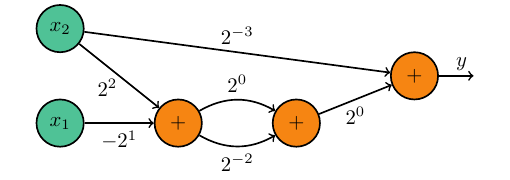}\label{fig::dagexample}
    }
    \vfil
    \subfloat[Pipelined \ac{dag}]{
    \includegraphics[width=.75\columnwidth]{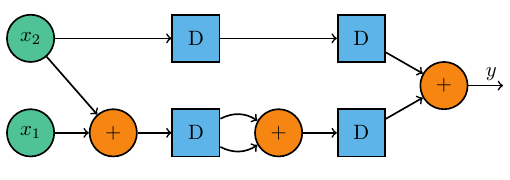}\label{fig::dagpipelining}
    }
    \caption{A \ac{dag} realizing the function $y(x_1,x_2)=(21/8)x_2-(5/4)x_1$ is depicted in (a). The same \ac{dag} is extended in (b) with delay elements to allow for pipelining.}
    \label{fig::pipelining}
\end{figure}
Three terms contribute to the overall computational cost
\begin{align}
    C_\mathrm{total} = C_\mathrm{add} N_\mathrm{add} + C_\mathrm{delay} N_\mathrm{delay} + C_\mathrm{inv} N_\mathrm{inv}.
\end{align}
The number of additions $N_\mathrm{add}$, the number of delay elements (latches) $N_\mathrm{delay}$ and the number of sign inverters $N_\mathrm{inv}$ required.
Further, $C_\mathrm{add}$, $C_\mathrm{delay}$ and $C_\mathrm{inv}$ are the effective cost for an addition, a delay element and an inverter, respectively.
Inspired by the CMOS implementation of these basic functions, we assume for simplicity that the cost for an adder and a delay element are approximately equal and set to\footnote{The cost of a full adder ranges around 20 transistors and can vary depending on the specific implementation used, clock speed, etc.
This cost only considers a full adder for the addition of two inputs of a single bit. For larger bitwidths the cost scales accordingly and simplifications in the implementation are possible. For simplicity we only consider the cost per bit.} $C_\mathrm{add} = C_\mathrm{delay} = 20$.
For an inverter we assume a cost of $C_\mathrm{inv} = 2$, since these can be easily implemented by two transistors~\cite{TietzeSchenk_2019}.

The number of additions in computing a \ac{dag} is upper bounded, as zeros are allowed for coefficients as well, by
\begin{subequations}
\begin{align}\label{eq::noadd}
    N_\mathrm{add} &= \sum_{i=K+1}^{|\mathcal{C}|} \left( \degi{\ve{c}_i} - 1 \right)  \\
    &\overset{\text{(a)}}{=} (|\mathcal{C}| - K) (S-1) \label{eq::noaddSfix}
\end{align}
\end{subequations}
where (a) follows from the fact that the number of additions $S-1$ for all vertices is constant. 

For medium to large matrices it may not be desirable to straightforwardly implement the \ac{dag} in hardware, apply a realisation of $\ve{x}$, and wait for the output $\ve{y}$ to be computed.
Particularly, for a \ac{dag} with many logical operations in sequence, this may take some time and is not an optimal use of resources.
Instead, a pipelined approach is desirable,\footnote{For a detailed discussion of pipelining, refer to~\cite{Hennessy_2012}.} each adder is followed by a latch or delay element that is able to store the intermediate result produced by that adder.
For example, after an addition is completed, and the result is stored, the following input realization can already be forwarded to the adder.
The stored result is then forwarded to the subsequent adder.
The schematic of a pipelined design is depicted in Fig.~\ref{fig::dagpipelining}.
There, a pipelined signal flow graph/\ac{dag} with two inputs $x_1$ and $x_2$ computes a single output $y$. 
The second input is required for the final addition computing the output.
Thus, two additional delay elements are required in the upper branch to delay the input accordingly, adding to the overall hardware cost.

Pipelining largely improves overall throughput, keeping each adder busy and reducing idle times of resources.
However, to enable that, idle paths require additional delay elements that contribute to the overall hardware cost.
Hence, for a practical algorithm it is desirable to not only minimize the number of adders but to find a \ac{dag} structure that limits the number of delay elements.
The overall number of delay elements required for a pipelined implementation of a decomposition \ac{dag} can be computed by
\begin{subequations}
\begin{align}
    N_\mathrm{delay} &= N_\mathrm{add} + \sum_{\forall \Tilde{\ve{c}} \in \Tilde{\mathcal{C}}} \; \left(\underset{
    \ve{c} \in \mathcal{D}(\Tilde{\ve{c}})}{\max} \mu_\mathrm{D} (\ve{c}) - \mu_\mathrm{D} (\Tilde{\ve{c}}) - 1 \right)
\end{align}
with
\begin{align}
    \Tilde{\mathcal{C}} &= \left\{ \ve{c} \in \mathcal{C} | \dego{\ve{c}} > 0 \right\} \\
    \mathcal{D}(\Tilde{\ve{c}}) &= \left\{\ve{c} \in \mathcal{C} | (\Tilde{\ve{c}}, \ve{c}) \in \mathcal{A} \right\}
\end{align}
\end{subequations}
The set $\mathcal{D}(\Tilde{\ve{c}})$ contains all vertices $\ve{c}$ that are connected by a directed arc in $\mathcal{A}$ from $\Tilde{\ve{c}}$ to $\ve{c}$.
The total number of delay elements is the sum of the number of adders, as each adder needs a buffer at the output, and for each node with outgoing arcs the longest path difference minus one that needs to be equalized.

The number of inverters depends on the specific algorithm used. 
For brevity, we will not discuss inverters in detail.
A reduction algorithm for the number of inverters in parallel \ac{lcc} algorithms is discussed in~\cite{Lehnert_2023:2}.

\section{Algorithmic Approaches}
We now discuss two existing algorithmic approaches for \ac{lcc}, namely a fully sequential and fully parallel algorithm.
Utilizing the best of both worlds, we introduce a new \ac{ua} that enables us to tune the \ac{dag} structure for further analysis.

\begin{figure*}[!t]
    \centering
    \subfloat[Fully Sequential Algorithm]{\includegraphics[width=1.5in]{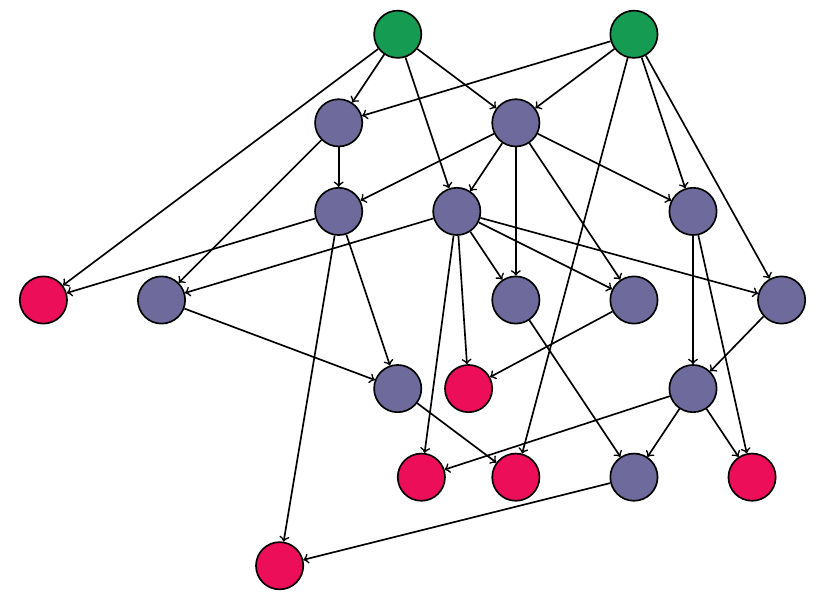}\label{fig::lz}}
    \hfil
    \subfloat[Fully Parallel Algorithm]{\includegraphics[width=1.5in]{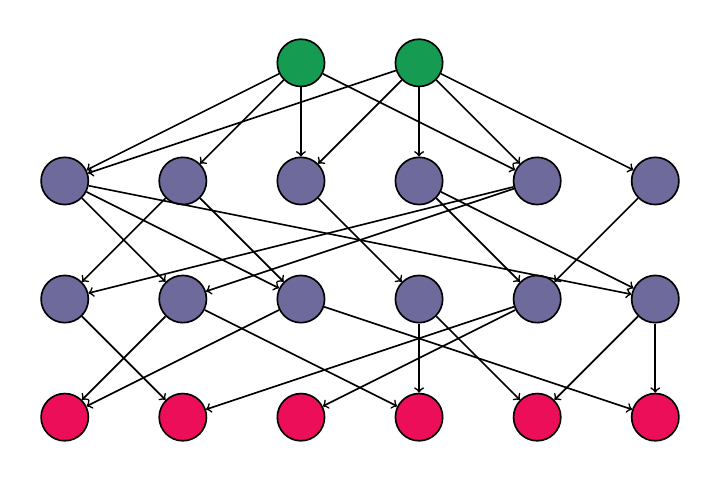}\label{fig::fp}}
    \hfil
    \subfloat[Mixed Alg. ($\Delta\mu_\mathrm{max}=0$)]{\includegraphics[width=1.5in]{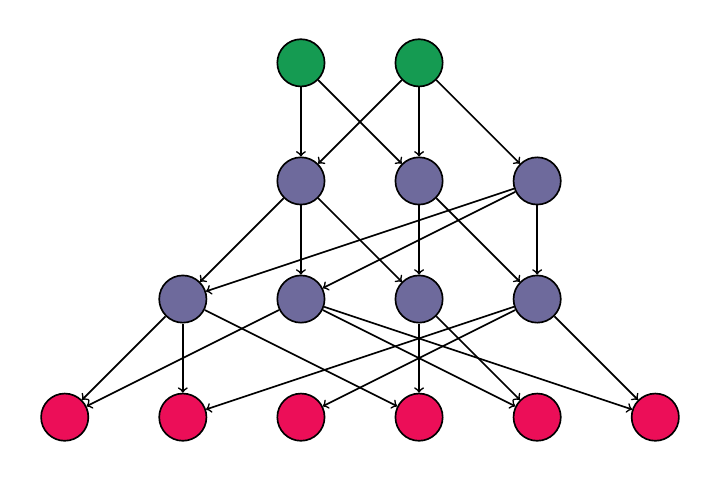}\label{fig::pa_fp}}
    \hfil
    \subfloat[Mixed Alg. ($\Delta\mu_\mathrm{max}=1$)]{\includegraphics[width=1.4in]{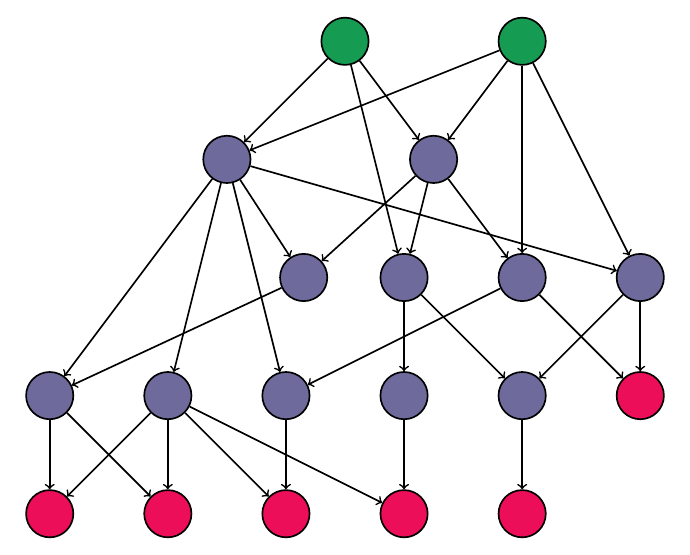}\label{fig::pa_l1}}
    \caption{Resulting graph topologies of different algorithmic approaches for decomposing a target matrix $\ma{T}$ of dimension $6 \times 2$. Green nodes represent input vertices corresponding to elements of the input vector $\ve{x}$, red nodes represent output vertices of the resulting matrix-vector product $\ve{y}$ and blue nodes are intermediary vertices of the decomposition graph.}
    \label{fig::topologies}
\end{figure*}

\subsection{\Acf{fs} Algorithm}
Given the set of all unit vectors in $K$ dimensions as our initial codebook set $\mathcal{C} = \left\{ \ve{1}_{1,K}, \dots, \ve{1}_{K,K} \right\}$, we recursively add vertices to the \ac{dag} using the following update rule~\cite{Muller_2022}:
\begin{align}\label{eq::updatefp}
    \mathcal{C} &\gets \mathcal{C} \cup \left\{ w(\ve{t}_{\Tilde{n}}, \mathcal{C}, S) \tomat{ \mathcal{C} } \right\}.
\end{align}
This means that we find the best linear combination of vectors in $\mathcal{C}$ that approximates $\ve{t}_{\Tilde{n}}$ well and requires $S-1$ additions.
We choose the row vector with index $\Tilde{n}$ from $\ma{T}$ that provides us with the largest reduction of the squared error for the update:
\begin{align}\label{eq::idxfp}
    \Tilde{n} = \underset{n \in \shortset{N}}{\argmin} \biggl( &\norm{ \ve{t}_n -  w(\ve{t}_n, \mathcal{C}, S) \tomat{ \mathcal{C} } }_2^2 + \nonumber\\ &\sum_{k \neq n} \norm{ \ve{t}_k - w(\ve{t}_k,\mathcal{C}, 1) \tomat{ \mathcal{C} } }_2^2 \biggr)
\end{align}
Although this approach shows excellent performance when looking at the tradeoff between distortion and the number of additions required, it is in many cases not suited for pipelining.
This follows from the fact that any $S$ vertices in a given codebook can be combined in each iteration, the obtained graph has an arbitrary structure (c.f. Fig.~\ref{fig::lz}).
Assuming for simplicity that each fundamental operation takes time $t_\mathrm{f}$ to compute,\footnote{This assumption is valid as long as we use the same type of adder throughout a \ac{dag}, i.e. $S$ is fixed.} it is concluded that the delay at any node $\ve{c}$ is $\mu_\mathrm{D}(\ve{c}) t_\mathrm{f}$.
Thus, if the depth $\mu_\mathrm{D}(\ve{c})$ varies in $\ve{c}$, delays are introduced that need to be compensated for.
The additional hardware resources and overhead required by the \ac{fs} algorithm are typically not acceptable, especially for large matrices. 
Therefore, algorithms that take these hardware constraints into account are desirable. 

\subsection{\Acf{fp} algorithm}
Instead of performing updates sequentially, we now successively refine the codebook for all vectors of the target matrix in parallel and then forget the old codebook.
Such a fully parallel algorithm can be written as a product of matrices~\cite{M_ller_2022}:
\begin{align}
    \ma{T} \approx \ma{W}_L \ma{W}_{L-1} \cdots \ma{W}_2 \ma{W}_1 \ma{C}_{0}.
\end{align}
The $n$-th row of the $l$-th matrix factor $\ma{W}_l$ is recursively obtained by
\begin{align}
    \ve{w}_{l,n} &= w(\ve{t}_n, \ma{C}_{l-1}, S) \quad \forall n \in [ N ]
\end{align}
with
\begin{align}
    \ma{C}_{l-1} &= \ma{W}_{l-1} \ma{W}_{l-2} \cdots \ma{W}_2 \ma{W}_{1} \ma{C}_0.
\end{align}
Each layer $l$ refines the approximation for each $\ve{t}_n$ using the codebook obtained in the previous iteration $l-1$.
Using our \ac{dag} based interpretation, this is the same as effectively restricting the codebook used in iteration $l$ to the subset of vectors in $\mathcal{C}$ at depth $l-1$, i.e. the matrix $\ma{C}_{l-1}$ contains all row vectors that are also included in the set $\{ \ve{c} \in \mathcal{C} | \mu_\mathrm{D}(\ve{c}) = l - 1 \}$.
As for the \ac{fs} algorithm, we use the set of all unit vectors as the initial codebook $\ma{C}_0 = \ma{I}_{N \times K}$.
The structure of the \ac{dag} generated by this algorithm is depicted in Fig.~\ref{fig::fp}.
Compared to the \ac{fs} algorithm, a fully parallel implementation in hardware can be achieved, no delays by differing path lengths are introduced.
However, this algorithm is not without drawbacks.
First, previous work~\cite{Rosenberger:2023} showed that refinement of the initial codebook during the first few iteration comes with a drop in performance.
Second, this algorithm does not scale to arbitrarily small matrices.
As the effective codebook scales with the target matrix size this can lead to convergence issues when decomposing smaller matrices.

\subsection{\Acf{ua}}
Using the ideas of the \ac{fs} and \ac{fp} algorithms we introduce a new \ac{ua} enabling us tune the structure of the computation \ac{dag}.
We reuse the sequential update rule from the \ac{fs} algorithm in~\eqref{eq::updatefp} to update $\mathcal{C}$.
\begin{subequations}
\begin{align}
    \Tilde{n} = \underset{n \in \shortset{N}}{\argmin} \; \lambda_n \biggl( &\norm{ \ve{t}_n -  w(\ve{t}_n, \mathcal{C}, S) \tomat{ \mathcal{C} } }_2^2 + \nonumber\\ &\sum_{k \neq n} \norm{ \ve{t}_k - w(\ve{t}_k,\mathcal{C}, 1) \tomat{ \mathcal{C} } }_2^2 \biggr)
\end{align}
with
\begin{align}
\lambda_n = \underset{j \in \mathcal{S} }{\max} \mu_\mathrm{D}(\ve{c}_j) \quad \text{and} \quad \mathcal{S} = \supp \left( w(\ve{t}_{n}, \mathcal{C}, S) \right). \nonumber
\end{align}
To obtain the index $\Tilde{n}$ for the target vector to be approximated, we amend the objective to update the approximation with the largest drop in error in~\eqref{eq::idxfp} by a multiplicative penalty factor $\lambda_n$.
This factor penalizes the absolute depth of approximations for different target vectors, i.e. updating a codeword at a higher depth leads to a larger penalty.
Moreover, to be able to limit the number and depth of idle paths in the \ac{dag}, we introduce a side constraint limiting the difference in depth for any linear combination of codewords, which is
\begin{align}
    &\underset{j \in \mathcal{S}}{\max}\left( \mu_\mathrm{D}(\ve{c}_j) \right) - \underset{j \in \mathcal{S}}{\min}\left( \mu_\mathrm{D}(\ve{c}_j) \right) \leq \Delta\mu_\mathrm{max}.
\end{align}
\end{subequations}
The parameter $\Delta\mu_\mathrm{max}$ controls the maximum difference in depth for the codewords used in each update.
For $\Delta\mu_\mathrm{max} \rightarrow \infty$ and $\lambda_n = 1$ the algorithm is equal to the \ac{fs} algorithm.
Constraining $\Delta\mu_\mathrm{max}=0$ we obtain a parallel structure of the decomposition \ac{dag}, similar to the \ac{fp} algorithm; however, codewords are added sequentially with a constraint on a parallel structure.
In general, the constraint on depth lets us tune the structure of the graph with respect to parallelism.
In Fig.~\ref{fig::pa_fp} and~\ref{fig::pa_l1}, the resulting graph structures for a graph constrained to a fully parallel structure and a depth difference of $\Delta\mu_\mathrm{max}=1$ are depicted, respectively.

\subsection{Related Algorithms} 
Most competing algorithms for \ac{cmvm} have a decent time complexity for small matrices.
However as they solve complex underlying problems, such as 0-1 integer linear programming~\cite{Aksoy_2015}, they do not scale well with growing matrix size and/or precision.
They are hence often intractable.
Instead, we use as a benchmark the best-performing \ac{mcm} algorithm known, presented in~\cite{Voronenko_2007}, that has reasonable polynomial time complexity and is thus tractable for larger matrices as well.
Note that \ac{mcm}, the multiplication of a variable scalar to a arbitrary constant vector, is a special case of \ac{cmvm}.
Any \ac{cmvm} problem can therefore be rewritten as a sum of $K$ \ac{mcm} problems, i.e.
\vspace{-3mm}
\begin{align}\label{eq::mcmslicing}
    \ve{y} = \ma{T}\ve{x} = \sum_{k=1}^K \ve{t}_k x_k,
\end{align}
that are solved independently.
Here, $\ve{t}_k$ and $x_k$ are the $k$-th column vector in $\ma{T}$ and $k$-th element in $\ve{x}$, respectively.
Due to the reduced search space the benchmark \ac{mcm} algorithm has excellent performance.
However, the adder tree required for the summation of the $K$ partial results, as well as a \ac{dag} structure, similar to the \ac{fs} algorithm, limit the performance when pipelined.

\section{Numerical Experiments}

\begin{figure}[!t]
    \centering
    \includegraphics[width=.9\columnwidth]{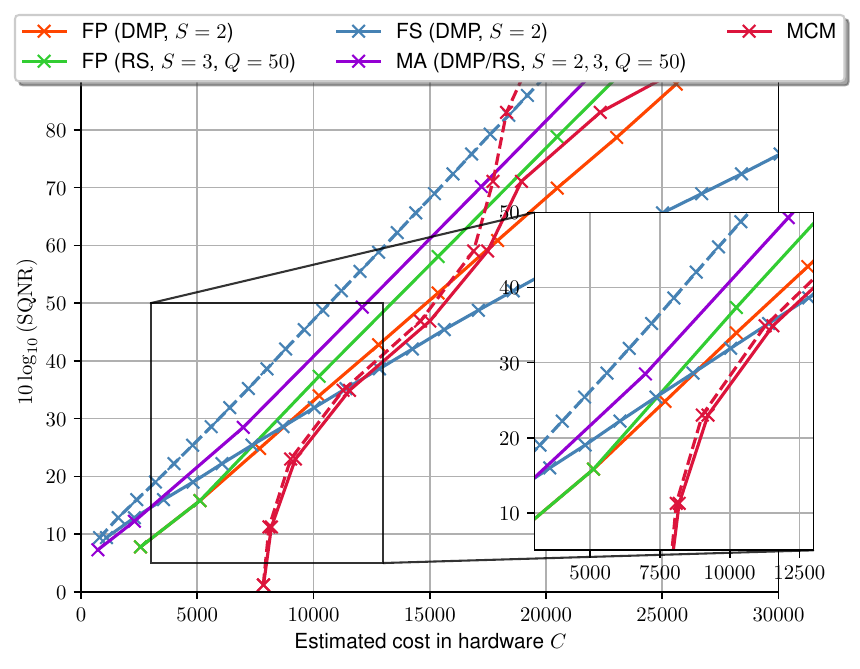}
    \caption{Comparison of different algorithmic approaches for decomposing a $64 \times 4$ target matrix $\ma{T}$. Solid lines indicate results considering the total cost $C_\mathrm{total}$. Dashed lines only consider the cost of adders $C_\mathrm{add} N_\mathrm{add}$. \ac{mcm} refers to the algorithm presented in~\cite{Voronenko_2007} (using the \textsc{C++} implementation available on~\cite{Spiral_2007} and extended by our hardware model). The results for each algorithm are averaged over $10^5$ matrix entries.}
    \label{fig::algcomp64x4}
\end{figure}

\begin{figure}[!t]
    \centering
    \includegraphics[width=.9\columnwidth]{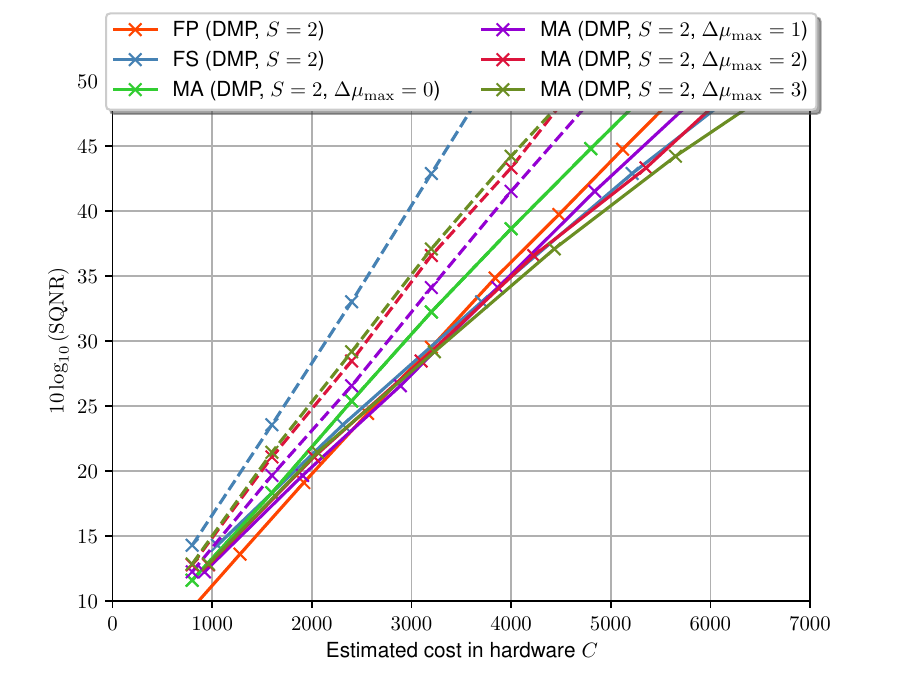}
    \caption{Comparison of different depth parameters $\Delta\mu_\mathrm{max}$ of the \ac{ua} given a $16 \times 4$ target matrix $\ma{T}$. Solid lines indicate results considering the total cost $C_\mathrm{total}$. Dashed lines only consider the cost of adders $C_\mathrm{add} N_\mathrm{add}$. The results for each algorithm are averaged over $10^5$ matrix entries.}
    \label{fig::algcompldiff16x4}
\end{figure}
The entries of all target matrices in the subsequent evaluations are drawn from an i.i.d. Gaussian distribution with zero mean and unit variance.
We expect that for practical matrices, e.g. weight matrices of \acp{nn}, similar performance is observed for \ac{lcc} algorithms~\cite{Mueller_2023}.
A Python implementation of all algorithms discussed in this paper is available in our \textit{github} repository: 
\url{https://github.com/hansrosenberger/computationcoding}.

As the first experiment, we compare the different algorithms for target matrices of dimension $64 \times 4$ in Fig.~\ref{fig::algcomp64x4}.
The figure shows, the \ac{fs} algorithm achieves the highest \ac{sqnr}, considering only the cost of additions (dashed lines).
However, when considering the total hardware cost, the \ac{fs} performance massively deteriorates, leaving this algorithm impractical for a pipelined implementation.
The overall hardware cost in this case is dominated by delay elements required to equalize path differences within the \ac{dag}.
The \ac{ua} constrained to a \ac{fp} structure shows the best overall performance, when considering the total hardware cost.
It outperforms the \ac{fp} algorithm, both the \ac{dmp} and \ac{rs} versions.
Relative gains are particulary large for the low \ac{sqnr} regime.
This is achieved by first setting $S=2$ and utilizing the \ac{dmp} to build up a coarse codebook from the initial codebook, and then dynamically switching to $S=3$ via the \ac{rs} approach.
The savings of \ac{ua} to the \ac{fs} result from an improved structure of the \ac{dag} for the first few layers.
The \ac{fp} algorithm is forced to find an approximation for each target vector separately.
This creates codewords that are correlated and unnecessary for the computation.
The \ac{ua} eliminates this redundancy (cf. Figs.~\ref{fig::fp} and~\ref{fig::pa_fp}).

As the second experiment we compare the performance of the \ac{ua} using different depth parameters $\Delta\mu_\mathrm{max}$ for target matrices of dimension $16 \times 4$ in Fig.~\ref{fig::algcompldiff16x4}.
Considering only the cost of the adders (dashed lines), we can clearly observe a tradeoff between parallelism and performance, i.e. decreasing $\Delta\mu_\mathrm{max}$ leads to a performance degradation.
However, when considering the total hardware cost (solid lines) the \ac{ua} performs best when constrained to a \ac{fp} structure ($\Delta\mu_\mathrm{max}=0$).
For $\Delta\mu_\mathrm{max}>0$ the \ac{ua} performs worse than its \ac{fp} counterpart and for some instances even worse than the \ac{fs} algorithm.
This result seems somewhat intuitive: Elements that incur a hardware cost that is not vanishingly small should also improve the \ac{sqnr}.
Hence, a fully parallel structure seems to be the best option.

\begin{remark}
\ac{lcc} works best for matrices with an exponential aspect ratio, i.e. $K \approx \log N$.
Therefore, we only consider in the evaluation matrices with that property.
For approximately square matrices it is beneficial to cut these into rectangular matrices with more extreme aspect ratios and apply an \ac{lcc} algorithm to each slice individually~\cite{Lehnert_2023}.
For example, to decompose a $64 \times 64$ matrix with a target \ac{sqnr} of \SI{47}{\decibel}, a slicing into submatrices of size $64 \times 4$ is a good choice. 

\section{Conclusion}
By interpreting the decomposition of a matrix as a \ac{dag}, we proposed a new \ac{ua} for \ac{lcc}.
The proposed algorithm is able to significantly outperform existing schemes.
Using a realistic hardware model for pipelining, we show that in almost all cases it is best to decompose a target matrix constraining the resulting \ac{dag} to a parallel structure.
\end{remark}

\bibliographystyle{ieeetr}
\bibliography{IEEEabrv,jabref_collection}

\end{document}